\documentclass[12pt]{iopart}
\usepackage{amstext}
\usepackage{amsbsy}
\usepackage{iopams}
\usepackage{graphicx}
\usepackage{color}
\begin{document}

\title[A stochastic analysis of the spatially extended May--Leonard model]
{A stochastic analysis of the spatially extended May--Leonard model}

\author{Shannon R. Serrao and Uwe C. T\"auber}

\address{Department of Physics (MC 0435) and 
		Center for Soft Matter and Biological Physics, Robeson Hall, 
		850 West Campus Drive, Virginia Tech, Blacksburg, VA 24061, USA}

\ead{shann87@vt.edu, tauber@vt.edu}

% submitted to: J. Phys. A: Math. Theor. (2017); arXiv: 1706.00309

\begin{abstract}
Numerical studies of the May--Leonard model for cyclically competing species 
exhibit spontaneous spatial structures in the form of spirals. It is desirable to 
obtain a simple coarse-grained evolution equation describing spatio-temporal 
pattern formation in such spatially extended stochastic population dynamics models. 
Extending earlier work on the corresponding deterministic system, we derive the 
complex Ginzburg--Landau equation as the effective representation of the fully 
stochastic dynamics of this paradigmatic model for cyclic dominance near its Hopf
bifurcation, and for small fluctuations in the three-species coexistence regime. 
The internal stochastic reaction noise is accounted for through the Doi--Peliti 
coherent-state path integral formalism, and subsequent mapping to three coupled 
non-linear Langevin equations. This analysis provides constraints on the model
parameters that allow time scale separation and in consequence a further reduction
to just two coarse-grained slow degrees of freedom.
\end{abstract}

\pacs{87.23.Cc, 02.50.Ey, 05.40.-a, 87.18.Tt}
% 87.23.Cc - Population dynamics and ecological pattern formation
% 02.50.Ey - Stochastic processes 
% 05.40.-a - Fluctuation phenomena, random processes, noise, and motion
% 87.18.Tt - Noise in biological systems

% \submitto{\JPA --- \today}

\section{Introduction} \label{sec:Introduction}

In the recent past, ecologists and applied mathematicians have sought to 
quantitatively delineate dynamic emerging phenomena such as biodiversity and  
population extinction~\cite{May73,Maynard74,Sigmund98,Murray02,Neal}. The payoff 
for these theoretical endeavors potentially includes stabilizing and protecting 
endangered ecosystems in addition to establishing a fundamental understanding of 
the myriad forms of pattern formation observed in nature. Populations with a cyclic
competition motif have been studied in several contexts such as ecology, 
epidemiology and opinion poll research. For example, cyclic dominance is observed
in certain Californian lizard subspecies~\cite{Lizard} and in petri dish
experiments that involve {\em E.-coli} bacteria with three distinct 
strains~\cite{Ecoli}. Stochastic spatially extended population dynamics with cyclic
motifs such as the May--Leonard model~\cite{MayLeo} have been investigated
numerically, mostly on two-dimensional lattices~\cite{0295-5075-117-4-48006,
17728757,He11}. In contrast to simpler cyclic rock-paper-scissors models for which 
the total particle number is conserved~\cite{Redner09,Mobilia10,He10}, May--Leonard
systems display striking spiral patterns in certain parameter ranges. Linking the 
characteristic length and time scales in these spontaneously emerging 
spatio-temporal structures to the basic rates of the underlying stochastic 
processes is an important fundamental problem. In the framework of the 
deterministic rate equation time evolution, Reichenbach, Mobilia, and 
Frey~\cite{Reichenbach2008368} demonstrated that the formation of spirals in the 
May--Leonard model is effectively governed by the paradigmatic complex 
Ginzburg--Landau equation (CGLE).

The CGLE appears extensively in various contexts in physics~\cite{cgle}, ranging
from second-order phase transitions in condensed matter systems to string theory 
and ubiquitous non-equilibrium phenomena (see, for example,
Refs.~\cite{Kuramoto,Cross,Cross09,Newell,Bohr,Dang,Pismen}). This complex partial 
differential equation typically describes a slowly varying continuous order 
parameter field in the presence of weak non-linearities near a bifurcation point 
that governs the instability of a spatially homogeneous state. It exhibits gauge 
invariance of the modulating variable under a global phase change, usually as a 
consequence of periodicity in the extended space-time. Intriguing more recent 
applications of the CGLE include the synchronization of coupled non-linear noisy 
oscillators~\cite{Risler05} and driven-dissipative Bose--Einstein condensation 
(through an equivalent Gross--Pitaevskii equation with complex 
parameters)~\cite{Sieberer13,Tauber14X,Liu16}, underscoring the remarkable 
universality of the CGLE.

More than thirty years ago, Kuramoto demonstrated that a lattice of diffusively 
coupled oscillators is governed by a generic coarse-grained evolution equation near
the Hopf bifurcation, namely the CGLE \cite{Kuramoto}. In 2007, Reichenbach, 
Mobilia, and Frey constructed the CGLE as a convenient effective description of the
May--Leonard model near the Hopf bifurcation of the three-species coexistence fixed 
point in parameter space~\cite{Reichenbach2008368}. They showed that the emerging 
spiral wavelength and wavefront velocity are encoded in the coefficients of the 
CGLE near this fixed point. In this present paper, we establish a full derivation
of the stochastic CGLE in this context, which properly accounts for intrinsic
reaction noise, and hence extends the deterministic analysis of 
Ref.~\cite{Reichenbach2008368}, and also the perturbative multi-scale expansion 
around the bifurcation performed by Szczesny, Mobilia, and 
Rucklidge~\cite{PhysRevE.90.032704}. We remark that the incorporation of intrinsic
stochasticity is crucial, as in some prominent situations, spatio-temporal patterns
cannot be adequately characterized by a mere deterministic treatment. This is true,
for example, in stochastic spatially extended lattice Lotka--Volterra models for
predator-prey competition and coexistence~\cite{Georgiev07,tauberLV,Chen16}. 
Under more general settings, Butler and Goldenfeld demonstrated that stochastic
fluctuations may cause significant alterations to otherwise simpler deterministic 
patterns~\cite{Goldenfeld09,Butler11}.

In our derivation of the stochastic CGLE for the May--Leonard model, valid near its
Hopf bifurcation and for small deviations from the stationary population densities 
in the three-species coexistence phase, we account for the systematic treatment of 
fluctuations due to internal reaction noise through a bosonic field-theoretic 
formalism. Thus, by including the inherent stochasticity of this system with cyclic
species competition as encoded in the microscopic master equation for its defining 
reaction processes, then studying (small) non-linear fluctuations about the 
mean-field stationary densities in the three-species coexistence region, and 
finally exploiting time scale separation in the vicinity of the Hopf bifurcation 
that allows us to eliminate one fast relaxing mode, we arrive at a Langevin-type 
extension of the complex Ginzburg--Landau equation, for which random effects are 
superimposed on the non-linear deterministic behavior ascribed to the CGLE through 
(to leading order) additive noise terms. Based on this resulting effective 
dynamical theory, one could now, e.g., in a perturbative analysis akin to 
Ref.~\cite{tauberLV}, evaluate fluctuation-induced renormalizations of the 
characteristic oscillation frequencies and attenuation, as well as typical spiral 
pattern wavelengths, and thereby quantitatively relate the stochastic CGLE 
components directly to numerical or actual observations in pattern formation. Yet
our formalism more also yields a set of three coupled Langevin equations that 
faithfully describes the stochastic spatially extended May--Leonard system under
quite general circumstances, not subject to the additional constraints required for
the applicability of the CGLE.

In the following, we first describe and define our spatially extended stochastic
version of the May--Leonard model, and provide the analysis of its relevant fixed 
points in section~\ref{MLmodelmf}. Next in section~\ref{DPpath} we derive the 
action describing the stochastic master equation evolution of the system through 
the Doi--Peliti coherent-state path integral formalism. We proceed by reducing the 
system to a coupled set of stochastic non-linear partial differential equations 
relevant to the three-species coexistence fixed point in section~\ref{Langevin}. 
In section~\ref{Invm}, we make crucial use of time scale separation near the Hopf 
bifurcation to obtain the invariant two-dimensional reactive manifold. On 
dimensional reduction, we recast the ensuing dynamical problem using normal forms, 
and derive the CGLE in section~\ref{norm} for small fluctuations within the species
coexistence phase. We explicitly articulate the noise contributions to the 
stochastic CGLE in section~\ref{noise}. In the concluding section~\ref{result}, we 
summarize our assumptions pertinent to this derivation, and comment on the range of
applicability of the CGLE mapping.

\section{Stochastic May-Leonard model} \label{MLmodelmf} 

\subsection{Model description}

The May--Leonard model for cyclic competition comprises the following independent 
stochastic processes
\begin{eqnarray}
  &X_{\alpha } + X_{\alpha + 1 } \to X_{\alpha } 
  \quad &{\rm with \ rate} \ \sigma' \ ,
  \label{Rcyclic} \\
  &X_{\alpha} \to X_{\alpha} + X_{\alpha} \quad &{\rm with \ rate} \ \mu \ .
  \label{Rrepro} 
\end{eqnarray}
Here, the subscript $\alpha = 1,2,3$ denotes the three competing populations, and
the identification $X_4 = X_1$ is implicit. Note that for simplicity we study the 
symmetric situation, for which identical reaction rates are implemented for all 
three species. 

In addition to these cyclic predation and reproduction reactions, we prescribe a 
population-limiting intra-species competition reaction,
\begin{equation}
  X_{\alpha} + X_{\alpha} \to X_{\alpha} \qquad\! {\rm with \ rate} \ \lambda' \ .
\label{Rdeath} 
\end{equation} 
We justify the addition of this reaction (which is not explicitly listed, e.g., in 
Ref.~\cite{Reichenbach2008368}) as follows: The May--Leonard model is usually 
simulated with population restrictions on the lattice (often, at most a single 
particle is allowed to occupy any site), representing finite local carrying
capacities $\rho$ for each species. For our subsequent theoretical analysis, it 
turns out that the pair coagulation \eref{Rdeath} represents a simpler `soft' 
implementation of local population density suppression than enforcing `hard' site 
restrictions; upon coarse-graining, both description become essentially equivalent 
(with an effective rate $\lambda'$ that can be expressed in terms of the 
reproduction rate $\mu$ and the local carrying capacity $\rho$)~\cite{tauberLV}.
Indeed, allowing multiple particles of either species to occupy each lattice site
will enable us to utilize bosonic field operators in section~\ref{DPpath} below.

In addition to the above on-site reactions, we allow for populations to migrate 
across the lattice through nearest-neighbor hopping (diffusion in the continuum
limit), 
\begin{equation}
  X_{\alpha , i}  \to X_{\alpha , i+1} \qquad\! {\rm with \ rate} \ D' \ ,
\label{RDiff} 
\end{equation} 
where the subscript $i$ denotes a lattice site (vector) index. The subsequent 
analysis of this stochastic spatially extended May--Leonard model variant will be 
carried out in the thermodynamic limit on an infinite lattice.

\subsection{Mean-field analysis}

We remark that our model represents a specialized case of generalized 
Lotka--Volterra systems, for which analyses of Hopf bifurcations and global 
Lyapunov functions are well-established, see, e.g., Ref.~\cite{Takeuchi}. 
In the (much simplified) case of a well-mixed system or for very fast diffusivity
$D' \gg \mu, \sigma', \lambda'$, spatial correlations are washed out and mean-field
mass action factorization is applicable. The corresponding coupled rate equations 
for the three spatially uniform particle densities $a_\alpha(t)$ read
\begin{equation}
  \frac{da_{\alpha + 1}(t)}{dt} = \mu \, a_{\alpha + 1}(t) 
  - \lambda \, a_{\alpha + 1}(t)^2 - \sigma \, a_\alpha(t) \, a_{\alpha + 1}(t) \ , 
\label{mfield}
\end{equation}
where we define the continuum reaction rates $\lambda = c^d \lambda'$ and
$\sigma = c^d \sigma'$, with $c$ denoting the lattice spacing. Mean-field steady
states are stationary solutions of the rate equations \eref{mfield}, 
$da_{\alpha}(t) / dt = 0$. There exist two sets of extinction fixed points, namely 
(i) $a_1 = \mu / \lambda$, $a_2 = 0 = a_3$ (and cyclic permutations thereof); and, 
provided $\lambda > \sigma$, (ii) $a_1 = \mu (\lambda - \sigma) / \lambda^2$, 
$a_2 = 0$, $a_3 = \mu / \lambda$ (and cyclic permutations thereof). Both sets of
extinction fixed points describe absorbing states in the stochastic system. In 
addition, we have a symmetric three-species coexistence fixed point 
$a_1 = a_2 = a_3 = \bar{a} = \mu / (\sigma + \lambda)$. 

\begin{figure}[t] \center 
\includegraphics[width=8cm]{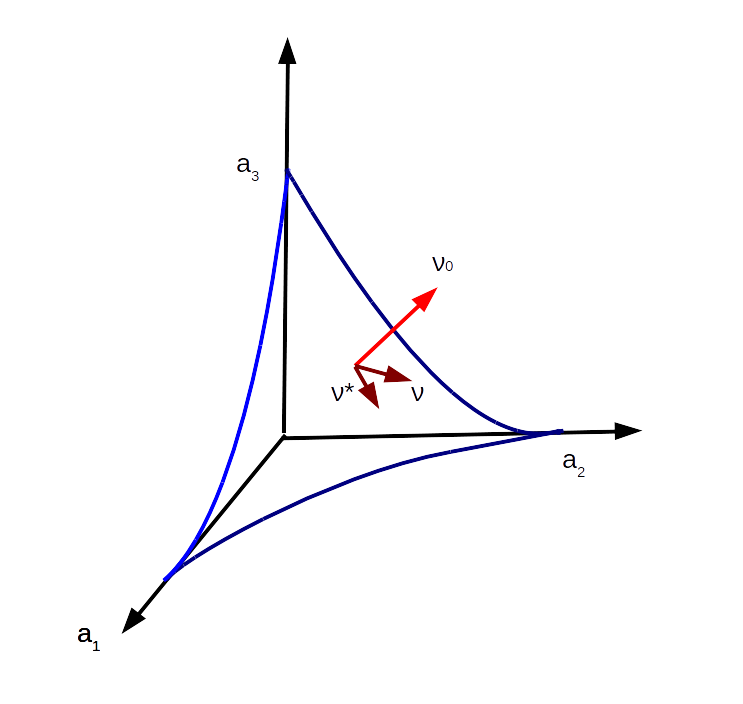}
\caption{The coexistence fixed point in the phase space of the three species
	densities $a_\alpha$ (in a non-spatial setting, or for a single lattice site). 
	The (short) brown arrows indicate the eigenvectors in the reactive manifold of 
	the system governed by damped oscillatory kinetics. The (long) red arrow 
	denotes the stable eigenvector; perturbations along this direction relax 
	exponentially towards the plane spanned by the two `slow' (near the Hopf 
	bifurcation) eigenvectors.} \label{concspace} 
\end{figure}

Linearizing about this coexistence fixed point, and collecting the species 
densities in a three-component vector $\boldsymbol{a} = (a_1,a_2,a_3)^T$, we have 
$\delta \dot{\boldsymbol{a}} = L \, \delta\boldsymbol{a}$, with the linear 
stability matrix 
\begin{equation}
  L = \frac{-\mu}{\sigma + \lambda} \ \left(  \begin{array}{ccc}
  \lambda & 0 & \sigma \\ \sigma  & \lambda & 0 \\ 0 & \sigma & \lambda \end{array}
  \right) \ . 
\end{equation}
Its eigenvalues at the coexistence fixed point are
\begin{equation}
  \nu_0 = -\mu \quad {\rm and} \quad 
  \lbrace \nu, \nu^* \rbrace = \frac{-\mu}{\sigma + \lambda} 
  \left( \lambda - \sigma \, e^{\mp i \pi / 3} \right) ; 
\label{evalue}
\end{equation}
the associated eigenvectors are depicted in Fig.~\ref{concspace}. The first 
eigenvalue $\nu_0 = - \mu$ is always negative, implying stability against small 
perturbations along its corresponding eigenvector, which will relax exponentially
in time $\sim e^{- \mu t}$ with decay rate $\mu$. The other two complex conjugate
eigenvalues describe either exponentially damped or growing temporal oscillations
with linear frequency $\omega_0 = \mu \sigma \sqrt{3} / 2 (\sigma + \lambda)$. For
$\lambda > \sigma / 2$, the real part of $\nu$ is negative, and the ensuing limit
cycle stable, contracting exponentially in time with decay rate 
$\mu (2 \lambda - \sigma) / 2 (\sigma + \lambda)$. Conversely, for 
$\sigma > 2 \lambda$ we obtain an unstable limit cycle, with an exponentially
growing amplitude. We note that in the associated spatially extended system, these
unstable limit cycles generate spiral structures, whose amplitudes are ultimately
constrained by the non-linear terms. 

In the language of bifurcation theory, when a pair of complex conjugate stability 
eigenvalues cross the imaginary axis of the complex plane, the associated dynamical
system displays a Hopf bifurcation; in our May--Leonard model variant it is located
at $\epsilon = (\sigma - 2 \lambda) / 2 (\sigma + \lambda) \to 0$. In the vicinity 
of the Hopf bifurcation, i.e., as the dimensionless parameter $|\epsilon| \ll 1$, 
the temporal evolution in the `reactive' plane spanned by the two eigenvectors 
corresponding to the eigenvalues $\nu$ and $\nu^*$ is very slow compared to the
fast relaxing mode along the orthogonal eigenvector corresponding to $\nu_0$. The
presence of the Hopf bifurcation thus provides us with a natural time scale 
separation for the dynamical eigenmodes of the May--Leonard system (see also 
Ref.~\cite{Labavic16}). The provision of a small expansion parameter $\epsilon$ 
distinguishes our model variant from the simpler one studied in 
Ref.~\cite{Reichenbach2008368}. The deterministic derivation of the CGLE carried 
out in Ref.~\cite{PhysRevE.90.032704} too utilizes time scale separation afforded 
through the small value of $\epsilon$ near the Hopf bifurcation.

\section{Doi--Peliti coherent-state path integral} \label{DPpath}

In order to systematically account for the intrinsic fluctuations in the system, 
we begin with the stochastic master equation. By applying the Doi--Peliti formalism
\cite{0305-4470-9-9-009,Peliti} (for more detailed pedagogical expositions, see
Refs.~\cite{Tauber05,cridyn}), we derive an effective field theory action that
captures the evolution of the system in the continuum limit, while faithfully 
incorporating its non-linearities and stochasticity due to the on-site reactions. 
We then construct an equivalent system of coupled Langevin equations which will 
serve as starting point to the subsequent derivation of the CGLE.

\subsection{Doi--Peliti operator representation of the master equation}

A specific configuration in this context entails enumerating the integer occupation
numbers $n_{\alpha , i} \geq 0$ for each species $\alpha$ at every lattice site $i$.
The state of the system at time $t$ is then given as a sum over all possible such
configurations, weighted with their probabilities $P(n_{\alpha , i};t)$ which change
over time through transitions with rates associated with the possible reactions 
allowed in the model. The continuous-time stochastic master equation describes the 
dynamical evolution of the system through balancing gain and loss terms for the
configurational probabilities. For the on-site stochastic reactions 
\eref{Rcyclic}--\eref{Rdeath}, excluding for now nearest-neighbor hopping, the
master equation reads explicitly:   
\begin{eqnarray}
  &&\frac{\partial P(n_{\alpha , i};t)}{\partial t} = \sum_{\alpha=1,2,3} \biggl( 
  \mu \, \Bigl[ (n_{\alpha , i}-1) \, P(n_{\alpha , i}-1;t) - n_{\alpha , i} \,
  P(n_{\alpha , i};t) \Bigr] \nonumber \\ 
  &&\quad\! + \sigma' \, \Bigl[ n_{\alpha , i}(n_{\alpha+1 , i}+1) \, 
  P(n_{\alpha , i} , n_{\alpha+1 , i}+1;t) - n_{\alpha , i} \, n_{\alpha+1 , i} \,
  P(n_{\alpha , i} , n_{\alpha+1 , i};t) \Bigr] \nonumber \\  
  &&\qquad\qquad\qquad\qquad + \lambda' \, \Bigl[ (n_{\alpha , i}+1) \, 
  P(n_{\alpha , i}+1;t) - n_{\alpha , i} \, P(n_{\alpha , i};t) \Bigr] \biggr) \ .
\end{eqnarray}
For the initial configuration, we assume the particle numbers on each site $i$ to 
be drawn from independent Poisson distributions with mean initial population 
densities $\bar{n}_\alpha = N_\alpha / N$, with the total number of lattice sites
$N$ and $N_{\alpha} = \sum_i n_{\alpha , i}$, i.e.:
\begin{equation}
  P(n_{\alpha , i};0) = \prod_{\alpha=1,2,3} 
  \frac{\bar{n}_{\alpha}^{n_{\alpha,i}}}{n_{\alpha,i}!} \, e^{- \bar{n}_\alpha} \ .
\end{equation}

We then adopt the ladder operator approach first associated with quantum harmonic 
oscillators to build up a many-particle Fock space on each site with the basic 
bosonic commutation relations $[a_{\alpha , i} , a_{\beta , j}] = 0 = 
[a^\dagger_{\alpha , i} , a^\dagger_{\beta , j}]$, 
$[a_{\alpha , i} , a^\dagger_{\beta , j}] = \delta_{ij} \delta_{\alpha\beta}$ and
particle number eigenstates $|n_{\alpha i} \rangle $ satisfying $a_{\alpha , i} \,
|n_{\alpha , i} \rangle = n_{\alpha , i} \, |n_{\alpha , i} - 1 \rangle$ and 
$a^\dagger_{\alpha , i} \, |n_{\alpha , i} \rangle = |n_{\alpha , i} + 1 \rangle$.
Any arbitrary state can then be written as a series of creation operators acting on
an empty vacuum state, $|n_{\alpha , i} \rangle = \prod_i \prod_\alpha 
(a^\dagger_{\alpha , i})^{n_{\alpha , i}} \, |0 \rangle$. A general state vector of 
the system $|\Phi(t) \rangle$ is then conveniently defined as the linear 
superposition of the various Fock space configurations of the system weighted by
their associated probabilities,   
\begin{equation}
  |\Phi (t) \rangle = \sum_{\lbrace n_{\alpha , i} \rbrace} P(n_{\alpha , i};t) \,
  |n_{\alpha , i} \rangle \ .
\end{equation}
The stochastic master equation can then be rewritten in this operator formalism as 
a time evolution operator $H$ acting on the state vector,
\begin{equation}
  \frac{\partial |\Phi (t) \rangle}{\partial t} = - H \, |\Phi(t) \rangle \ , 
\end{equation}
where the `pseudo-Hamiltonian' is a sum of products of (normal-ordered) creation 
and annihilation operators.

For our stochastic May--Leonard problem, the evolution operator $H_{\rm reac}$ 
corresponding to the local on-site reactions becomes
\begin{eqnarray}
  H_{\rm reac} &=& \sum_i^N \sum_{\alpha=1,2,3} \Bigl[ 
  \mu \, (1 - a^\dagger_{\alpha , i}) \, a^\dagger_{\alpha , i} a_{\alpha , i} 
  + \sigma'\, (a^\dagger_{\alpha+1 , i} - 1) \, a^\dagger_{\alpha , i} 
  a_{\alpha , i} a_{\alpha+1 , i} \nonumber \\
  &&\qquad\qquad + \lambda' \, (a^\dagger_{\alpha , i} - 1) \, 
  a^\dagger_{\alpha , i} a_{\alpha , i}^2 \Bigr] \ .
\end{eqnarray}
We may also construct the non-local pseudo-Hamiltonian $H_{\rm diff}$ describing
hopping transport on the lattice or unbiased diffusion with continuum diffusivity 
$D = c^2 D'$: 
\begin{equation}
  H_{\rm diff} = \sum_{\langle i,j \rangle} \sum_{\alpha=1,2,3} D' \, 
  (a^\dagger_{\alpha , i} - a^\dagger_{\alpha , j}) \, 
  (a_{\alpha , i} - a_{\alpha , j}) \ , 
\end{equation}
where we sum over nearest-neighbor pairs $\langle i , j \rangle$. We then express 
the total pseudo-Hamiltonian of the system as a sum of both these contributions,
$H = H_{\rm reac} + H_{\rm diff}$.

\subsection{Field theory action in the coherent-state basis}

Following the procedures detailed in Refs.~\cite{Tauber05,cridyn}, one can compute 
the expectation values of any observable 
$\mathcal{O}(\lbrace n_{\alpha , i} \rbrace)$ as a path integral over a 
coherent-state basis,
\begin{eqnarray}
  \langle \mathcal{O}\rangle \propto \int \prod_i^N \prod_{\alpha=1,2,3} 
  \mathcal{D}[\psi_{\alpha , i}^* , \psi_{\alpha , i}] \,
  \mathcal{O}(\lbrace \psi_{\alpha , i} \rbrace) \, 
  e^{- S[\psi_{\alpha , i}^* , \psi_{\alpha , i};t]} \ ,
\end{eqnarray}
the $\psi_{\alpha , i}^*$ and $\psi_{\alpha , i}$ respectively denoting the 
complex-valued left and right eigenvalues of $a^\dagger_{\alpha , i}$ and 
$a_{\alpha , i}$. The associated Doi--Peliti action for local on-site reactions is 
\begin{eqnarray}
  S[\psi_{\alpha , i}^* , \psi_{\alpha , i};t] &=& \int_0^{t_f} dt \, \biggr[ 
  \sum_i \sum_{\alpha=1,2,3} \psi_{\alpha , i}^*(t) \, 
  \frac{\partial \psi_{\alpha , i}(t)}{\partial t}
  + H(\psi_{\alpha , i}^*(t) , \psi_{\alpha , i}(t)) \biggr] \nonumber \\
  &&\qquad\qquad - \sum_i \sum_{\alpha=1,2,3} \Bigl[ \psi_{\alpha , i}(t_f) + 
  \bar{n}_{\alpha} \, \psi_{\alpha , i}^*(0) \Bigr] \ .
\end{eqnarray} 
In this expression, the last term originates from the initial Poissonian product
distribution, while the penultimate term corresponds to the field computed at the 
final time. The terms in square brackets, referred to as the `bulk' part of the 
action, are relevant to our subsequent analysis and derivation. The 
$H(\psi_{\alpha , i}^*(t) , \psi_{\alpha , i}(t))$ term in the bulk action is 
simply the evolution operator $H$, obtained by replacing the creation and
annihilation operators with the corresponding coherent-state eigenvalue fields.

Finally, we proceed to the continuum limit of our problem (lattice constant 
$c \to 0$) by substituting $\sum_{i=1}^N \to c^{-d} \int d^dx$, 
$\psi_{\alpha , i}(t) \to c^d a_\alpha(\vec{x},t)$, and 
$\psi_{\alpha , i}^*(t) \to 1 + \tilde{a}_\alpha(\vec{x},t)$, and thus obtain the
coarse-grained bulk action for the May--Leonard model, 
\begin{equation}
  S[\tilde{a}_\alpha , a_\alpha ; t] = \int dt \int d^dx \, \Biggl[ \
  \sum_{\alpha=1,2,3} \tilde{a}_\alpha (\partial_t - D \nabla^2) \, a_\alpha 
  + H_{\rm reac}(\tilde{a}_\alpha , a_\alpha) \Biggr] \ . 
\label{contaction}
\end{equation} 
Here, the continuum pseudo-Hamiltonian for the on-site reactions reads explicitly
\begin{eqnarray}
  H_{\rm reac}(\tilde{a}_\alpha , a_\alpha) &=& \sum_\alpha \Bigl[ 
  - \mu \, \tilde{a}_\alpha (\tilde{a}_\alpha + 1) \, a_\alpha + \sigma \, 
  \tilde{a}_{\alpha + 1} (\tilde{a}_\alpha + 1) \, a_\alpha a_{\alpha + 1} 
  \nonumber \\
  &&\qquad\! + \lambda \, \tilde{a}_\alpha (\tilde{a}_\alpha + 1) \, a_\alpha^2 
  \Bigr] \ ,
\label{shreac}  
\end{eqnarray}
with all contributions written in terms of the continuum reaction rates 
$\sigma = c^d \sigma'$, $\lambda = c^d \lambda'$, and $D = c^2 D'$.

\section{Langevin description} \label{Langevin}

In order to derive the CGLE, we seek a set of coupled stochastic partial 
differential equations
\begin{equation}
  \frac{\partial {\bf a}(\vec{x},t)}{\partial t} = D \nabla^2 {\bf a}(\vec{x},t) 
  + {\bf F}[{\bf a}(\vec{x},t)] + \boldsymbol{\zeta}(\vec{x},t) \ ,
\end{equation}
with $\langle \boldsymbol{\zeta} \rangle = 0$ and associated noise correlations
\begin{equation}
  \langle \zeta_\alpha(\vec{x},t) \, \zeta_\beta(\vec{x\,}',t') \rangle = 2
  L_{\alpha \beta}[{\bf a}(\vec{x},t)]\, \delta(\vec{x}-\vec{x\,}') \delta(t-t')\ .  
\end{equation}
A set of coupled stochastic partial differential equations of this form can be cast
in terms of an equivalent dynamical Janssen~de-Dominicis response functional
\cite{Janssen76,DeDominicis76,cridyn}  
\begin{eqnarray}
  S[{\bf a}] = \int dt \int d^dx \sum_\alpha \tilde{a}_\alpha \biggr[
  (\partial_t - D \nabla^2) \, a_\alpha - F_\alpha[{\bf a}] - \sum_\beta 
  L_{\alpha \beta}[{\bf a}] \, \tilde{a}_\beta \biggr] \ . 
\label{respf}  
\end{eqnarray}

Hence, upon identifying the response functional \eref{respf} with the bulk 
Doi--Peliti action in eqs.~\eref{contaction} and \eref{shreac}, one arrives at
coupled Langevin equations  
\begin{eqnarray}
  \frac{\partial a_{\alpha + 1}(\vec{x},t)}{\partial t} &=& (\mu + D \nabla^2) \,
   a_{\alpha + 1}(\vec{x},t) - \lambda \, a_{\alpha + 1}(\vec{x},t)^2 - \sigma \, 
   a_\alpha(\vec{x},t) \, a_{\alpha + 1}(\vec{x},t) \nonumber \\
   &&+ \zeta_{\alpha + 1}(\vec{x},t)
\end{eqnarray}
for the three complex fields $a_\alpha(\vec{x},t)$.\footnote{The identification of 
the `shifted' Doi--Peliti action with a dynamical response functional is associated 
with certain mathematical subtleties; for an up-to-date exposition and analysis, 
see Ref.~\cite{laneq16}.} On comparison with the mean-field rate equations 
\eref{mfield}, we note the presence of additional diffusion and multiplicative 
noise contributions that are governed by the (symmetric) stochastic correlation 
matrix
\begin{eqnarray}
  L_{\alpha \beta}[{\bf a}] = \left( \begin{array}{ccc} 
  \mu \, a_1 - \lambda \, a_1^2 & - \sigma \, a_1 a_2 / 2 & - \sigma \, a_1 a_3 / 2 
  \\ 
  - \sigma \, a_1 a_2 / 2 & \mu \, a_2 - \lambda \, a_2^2 & - \sigma \, a_2 a_3 / 2 
  \\ 
  - \sigma \, a_1 a_3 / 2 & - \sigma \, a_2 a_3 / 2 & \mu \, a_3 - \lambda \, a_3^2 
  \end{array} \right) .
\end{eqnarray}

We proceed with a linear variable transformation to fluctuating dynamical fields
relative to the three-species (mean-field) coexistence fixed-point densities
$\bar{a} = \mu / (\sigma + \lambda)$,
\begin{equation}
  a_\alpha(\vec{x},t) = \frac{\mu}{\sigma + \lambda} + b_{\alpha}(\vec{x},t) \ , 
  \qquad \tilde{a}_\alpha(\vec{x},t) = \tilde{b}_\alpha(\vec{x},t) \ .
\end{equation} 
The $b$ fields thus represent deviations about the mean-field stationary
concentrations, governed by stochastic partial differential equations of the form
\begin{equation}
  \frac{\partial {\bf b}(\vec{x},t)}{\partial t} = D \nabla^2 {\bf b}(\vec{x},t) 
  + {\bf f}[{\bf b}(\vec{x},t)] + \boldsymbol{\zeta'}(\vec{x},t) \ ,
\end{equation}
with ${\bf f}[b_\alpha] = {\bf F}[a_\alpha \to \bar{a} + b_\alpha]$ and 
$\langle \zeta'_\alpha \, \zeta'_\beta \rangle = 2 B_{\alpha \beta} \,
\delta(\vec{x}-\vec{x\,}') \delta(t-t')$, 
$B_{\alpha \beta} = L_{\alpha \beta}[a_\alpha \to \bar{a} + b_\alpha]$.

\section{Invariant manifold} \label{Invm}

Our main goal is to explore if the stochastic May--Leonard problem can possibly be
further simplified by reducing the dynamical degrees of freedom from three to two,
at least for sufficiently small fluctuations in the species coexistence regime. To 
this end, we essentially follow the procedure for the deterministic model outlined
in Ref.~\cite{Reichenbach2008368}, and utilize its invariant manifold, i.e., the 
dynamical subspace of the system invariant to perturbations. Within the linear 
approximation, the deterministic dynamics will stabilize on the reactive plane 
spanned by the eigenvectors associated with the complex eigenvalues $\nu$ and
$\nu^*$ normal to the eigendirection with negative eigenvalue $\nu_0 = - \mu$. 
Near the Hopf bifurcation, this stable mode relaxes very fast in comparison with 
the oscillations in the invariant manifold and can hence be eliminated. We shall
first apply a suitable linear variable transformation that allows us to orient the 
dynamical degrees of freedom along the stable (fast) and reactive (slow) 
directions. Note that the dynamics of the (to linear order) fast relaxing mode will 
be affected by non-linear couplings as well as noise cross-correlations to the two
oscillating modes, and in turn feed back into the slow dynamics on the invariant
manifold. Consequently, as a second step, we shall exploit time scale separation 
afforded by the vicinity to the Hopf bifurcation and slave the fast mode to the two
slow degrees of freedom and thereby account for these non-linear effects. We remark
that while we shall carry out the analysis in the Langevin representation here, we
could have equivalently performed all required transformations within the 
associated dynamical functionals.
  
\subsection{Dynamical variable transformation} \label{dvt}

We proceed with a non-orthogonal dynamical variable transformation but otherwise
akin to a rotation aligning the dynamical variables within and normal to the 
invariant manifold. Introducing the column vector ${\bf c} =(c_1, c_2, c_3)^T$ for
the new fields, we apply a transformation $\bf{c}= \mathcal{R} \bf{b}$, where 
\begin{equation}
  \mathcal{R}  = \left( \begin{array}{rcr} 1 / \sqrt{2} & 0 & - 1 / \sqrt{2} \\
  - 1 / \sqrt{6} & \sqrt{2/3} & - 1 / \sqrt{6} \\
  1 / \sqrt{3} & 1 / \sqrt{3} & 1 / \sqrt{3} \end{array} \right) .
\end{equation}
Following this transformation, we obtain a set of stochastic partial differential 
equations in the new $c$ fields \cite{1751-8121-46-29-295002},
\begin{equation}
  \partial_t {\bf c} = D \nabla^2 {\bf c} + \mathcal{R} f[\mathcal{R}^{-1}{\bf c}] 
  + \boldsymbol{\eta} 
\end{equation}
where $\langle \eta_\alpha \, \eta_\beta \rangle = 2 \tilde{B}_{\alpha \beta} \,
\delta(\vec{x}-\vec{x\,}') \delta(t-t')$, $\tilde{B}_{\alpha \beta} = 
[\mathcal{R} B \mathcal{R}^T]_{\alpha \beta}$ ($\alpha,\beta=1,2,3$). Explicitly,
the resulting coupled Langevin equations become
\begin{eqnarray} 
  \partial_t c_1 &=& D \nabla^2 c_1 
  + \frac{\mu (\sigma - 2 \lambda)}{2 (\sigma + \lambda)} \, c_1 
  + \frac{\sqrt{3} \mu \sigma}{2 (\sigma + \lambda)} \, c_2 
  + \frac{\sigma}{2 \sqrt{2}} \, c_1^2 - \frac{\sigma}{2 \sqrt{2}} \, c_2^2 
  \nonumber \\ 
  &&- \frac{\sigma - 2 \lambda}{\sqrt{6}} \, c_1 c_2 
  - \frac{\sigma + 4 \lambda}{2 \sqrt{3}} \, c_1 c_3  
  + \frac{\sigma}{2} \, c_2 c_3 + \eta_1 \ , 
\label{1ceq} \\
  \partial_t c_2 &=& D \nabla^2 c_2 
  + \frac{\mu (\sigma - 2 \lambda)}{2 (\sigma + \lambda)} \, c_2
  - \frac{\sqrt{3} \mu \sigma}{2 (\sigma + \lambda)} \, c_1
  - \frac{\sigma - 2 \lambda}{2 \sqrt{6}} \, c_1^2 
  + \frac{\sigma - 2 \lambda}{2 \sqrt{6}} \, c_2^2 \nonumber \\ 
  &&- \frac{\sigma}{\sqrt{2}} \, c_1 c_2 
  - \frac{\sigma + 4 \lambda}{2 \sqrt{3}} \, c_2 c_3
  - \frac{\sigma}{2} \, c_1 c_3 + \eta_2 \ , 
\label{2ceq} \\
  \partial_t c_3 &=& D \nabla^2 c_3 - \mu \, c_3 
  + \frac{\sigma - 2 \lambda}{2 \sqrt{3}} \, c_1^2 
  + \frac{\sigma - 2 \lambda}{2 \sqrt{3}} \, c_2^2
  - \frac{\sigma + \lambda}{\sqrt{3}} \, c_3^2 + \eta_3 \ .
\label{3ceq}
\end{eqnarray}
Inspection of eq.~\eref{3ceq} confirms that the linear couplings of $c_3$ to the 
other modes have indeed been eliminated. Note that the `mass' terms, i.e., constant
coefficients of the $c_1$ and $c_2$ terms in eqs.~\eref{1ceq} and \eref{2ceq}, 
respectively, are $\mu \epsilon$. When compared to the corresponding relaxation 
rate $\mu$ in the term linear in $c_3$ in eq.~\eref{3ceq}, we see again that as 
$\epsilon \to 0$, $c_3$ relaxes much faster than $c_1$ and $c_2$ (to linear order).
This time scale separation provides the rationale for the subsequent elimination of
the $c_3$ degree of freedom. Yet if the dimensionless parameter $\epsilon$ is not 
small, no such reduction to just two dynamical degrees of freedom can be justified,
and hence a mapping to the CGLE is inapplicable. One then needs to retain all three
dynamical modes and their non-linear couplings and noise (cross-)correlations to 
faithfully describe the dynamics of the system. The effective description of the
May--Leonard model in terms of the CGLE consequently holds only near the Hopf
bifurcation. 

Let us further consider the linear terms in the equations of motion for the `slow'
modes $c_1$ and $c_2$. We see that for $\epsilon < 0$, i.e., $\sigma < 2 \lambda$,
deviations from the coexistence fixed point densities will exponentially relax to
zero. Comparison with the diffusive spreading term yields a characteristic length
scale $\xi_c = \sqrt{\frac{2D \, (\lambda + \sigma)}{\mu \, (2 \lambda - \sigma)}}$ 
which describes the typical extent of spatial patches for each species. One would
not expect to encounter other more interesting spatio-temporal structures in this
regime. On the other hand, for $\epsilon > 0$ or $\sigma > 2 \lambda$, the 
coexistence fixed point becomes unstable, and correspondingly, spatially 
homogeneous species distributions develop instabilities at wavelengths larger than
$\lambda_c = \sqrt{\frac{2D \, (\sigma + \lambda)}{\mu \, (\sigma - 2 \lambda)}}$. 
Along with the associated periodic temporal oscillations, this generates spiral
structures of typical size $\lambda_c$. Note that these are stabilized due to 
saturating non-linearities in the fluctuating fields $c_1$ and $c_2$, which we 
capture in the next section. 

\subsection{Dimensional reduction} \label{drim}

We seek to describe the invariant manifold through a function 
$c_3 = \mathcal{G}(c_1, c_2)$ that expresses the fast variable in our problem 
through the two slow modes. It is a difficult problem to obtain $\mathcal{G}$ 
exactly to all orders. However, one can try the simplest non-trivial ansatz 
compatible with the rotational symmetry in the reactive plane spanned by $c_1$ and
$c_2$; to first approximation, we set  
\begin{equation} 
  c_3 = K \left( c_1^2 + c_2^2 \right) ,
\label{slave}
\end{equation}
where $K$ is a constant, to be determined next. Differentiation with respect to 
time gives $\partial_t c_3 = 2 K \, (c_1 \partial_t c_1 + c_2 \partial_t c_2)$. We 
then just substitute the deterministic part of eqs.~\eref{1ceq}-\eref{3ceq} to 
identify $K$ to second order in $c_1$ and $c_2$. To simplify the calculation 
involving the Laplacian operators, we operate in spatial Fourier space, replacing
the $\nabla^2$ operators with $- p^2$. This yields
\begin{equation}
  K = \frac{(\sigma - 2 \lambda ) (\sigma + \lambda)}
  {2 \sqrt{3} \, \mu (2 \sigma - \lambda)} \left( 1 - \frac{D (\sigma + \lambda)}
  {\mu (2 \sigma - \lambda)} \, p^2 \right)^{-1} . 
\label{Kcond}  
\end{equation}
Near the Hopf bifurcation, where $\sigma \approx 2 \lambda$ and in the 
long-wavelength limit $p \ll \sqrt{\mu / D}$, the inverse bracket in \eref{Kcond} 
approximately becomes $1 + D p^2 / \mu$, whence we may set
\begin{equation} \label{Kvalue}
  K \approx \frac{(\sigma - 2 \lambda) (\sigma + \lambda)}
  {2 \sqrt{3} \, \mu (2 \sigma - \lambda)} \ , 
\end{equation}
and subsequently restore all $p^2$ terms with the differential operators 
$-\nabla^2$. Inserting this result \eref{Kvalue} into eqs.~\eref{1ceq}, \eref{2ceq}
we obtain
\begin{eqnarray} 
  &&\partial_t c_1 = D \nabla^2 c_1
  + \frac{\mu (\sigma - 2 \lambda)}{2 (\sigma + \lambda)} \, c_1
  + \frac{\sqrt{3} \mu \sigma}{2 (\sigma + \lambda)} \, c_2 
  + \frac{\sigma }{2 \sqrt{2}} \, c_1^2
  - \frac{\sigma }{2 \sqrt{2}} \, c_2^2 \nonumber \\ 
  &&\quad\;\ - \frac{\sigma - 2 \lambda}{\sqrt{6}} \, c_1 c_2 
  - \frac{(\sigma - 2 \lambda) (\sigma + \lambda) (\sigma + 4 \lambda)}
  {12 \mu \, (2 \sigma - \lambda)} \, c_1^3 
  + \frac{\sigma (\sigma - 2 \lambda) (\sigma + \lambda)}
  {4 \sqrt{3} \mu \, (2 \sigma - \lambda)} \, c_1^2 c_2 \nonumber \\ 
  &&\qquad - \frac{(\sigma - 2 \lambda) (\sigma + \lambda) (\sigma + 4 \lambda)}
  {12 \mu \, (2 \sigma - \lambda)} \, c_1 c_2^2 
  + \frac{\sigma (\sigma - 2 \lambda) (\sigma + \lambda)}
  {4 \sqrt{3} \mu \, (2 \sigma - \lambda)} \, c_2^3 + \eta_1 \ , 
\label{1fceq} \\
  &&\partial_t c_2 = D \nabla^2 c_2 
  + \frac{\mu (\sigma - 2 \lambda)}{2 (\sigma + \lambda)} \, c_2
  - \frac{\sqrt{3} \mu \sigma}{2 (\sigma + \lambda)} \, c_1
  - \frac{\sigma - 2 \lambda}{2 \sqrt{6}} \, c_1^2 
  + \frac{\sigma - 2 \lambda}{2 \sqrt{6}} \, c_2^2 \nonumber \\ 
  &&\qquad - \frac{\sigma}{\sqrt{2}} \, c_1 c_2 
  - \frac{\sigma (\sigma - 2 \lambda) (\sigma + \lambda)}
  {4 \sqrt{3} \mu\, (2 \sigma - \lambda)} \, c_1^3 
  - \frac{(\sigma - 2 \lambda) (\sigma + \lambda) (\sigma + 4 \lambda)}
  {12 \mu \, (2 \sigma - \lambda)} \, c_1^2 c_2 \nonumber \\ 
  &&\qquad - \frac{\sigma (\sigma -2 \lambda ) (\sigma + \lambda)}
  {4 \sqrt{3} \mu \, (2 \sigma - \lambda)} \, c_1 c_2^2 
  - \frac{(\sigma - 2 \lambda) (\sigma + \lambda) (\sigma + 4 \lambda)}
  {12 \mu \, (2 \sigma - \lambda)} \, c_2^3 + \eta_2 \ .
\label{2fceq}
\end{eqnarray}
The wavevector-dependent contributions from \eref{Kcond} would induce additional
subleading non-linearities containing spatial derivatives. In the noise covariance
matrix we also simply apply the substitution \eref{Kvalue}, as detailed in 
section~\ref{noise} below.

\section{Derivation of the CGLE from the normal form} \label{norm}

The normal form of a dynamical system encapsulates its essential behavior. Normal
forms facilitate the description of non-linear dynamics near bifurcations in a 
natural way, and thus enable classification 
schemes~\cite{Wiggins90,Arrowsmith,Guckenheimer,Murdock}. Reichenbach, Mobilia, and 
Frey showed that the normal form of the deterministic May--Leonard model allows its
characterization in terms of the CGLE as its effective dynamical 
description~\cite{Reichenbach2008368}. We therefore proceed to obtain a non-linear 
variable transformation $c_i \to z_i$ with the goal to eliminate the quadratic 
terms in eqs.~\eref{1fceq} and \eref{2fceq}:
\begin{eqnarray}
  z_1 &=& c_1 + \frac{1}{\mu (7 \sigma^2 - \sigma \lambda + \lambda^2)} \Biggr[
  \frac{\sigma (\sigma - 2 \lambda) (\sigma + \lambda)}{2 \sqrt{2}} \, c_1^2 
  \nonumber \\ 
  &&\qquad\qquad + \frac{5 \sigma^3 + 3 \sigma^2 \lambda + 2 \lambda^3}{\sqrt{6}} 
  \, c_1 c_2 - \frac{\sigma (\sigma - 2 \lambda) (\sigma + \lambda)}{2 \sqrt{2}} 
  \, c_2^2 \Biggr] \ , 
\label{ntz1} \\
  z_2 &=& c_2 + \frac{1}{\mu (7 \sigma^2 - \sigma \lambda + \lambda^2)} \Biggr[
  \frac{5 \sigma^3 + 3 \sigma^2 \lambda + 2 \lambda^3}{2 \sqrt{6}} \, c_1^2 
  \nonumber \\ 
  &&\qquad\qquad - \frac{\sigma (\sigma - 2 \lambda) (\sigma + \lambda)}{\sqrt{2}} 
  \, c_1 c_2 - \frac{5 \sigma^3 + 3 \sigma^2 \lambda + 2 \lambda^3}{2 \sqrt{6}} 
  \, c_2^2 \Biggr] \ .
\label{ntz2}
\end{eqnarray}
 
This non-linear transformation enables us to write the effective dynamical system 
in the following form, up to quartic terms in the complex fields 
${\bf z} = (z_1,z_2)$:
\begin{eqnarray}
  \partial_t z_1 &=& D \nabla^2 z_1 + k_1 z_1 + k_2 z_2 - k_3 (z_1 + k_4 z_2)
  (z_1^2 + z_2^2) + \mathcal{O}({\bf z}^4) + \eta_1 \ , \nonumber \\
  \partial_t z_2 &=& D \nabla^2 z_2 + k_1 z_2 - k_2 z_1 - k_3 (z_2 - k_4 z_1)
  (z_1^2 + z_2^2) + \mathcal{O}({\bf z}^4) + \eta_2 \ , 
\label{gclecomp}  
\end{eqnarray}
with coefficients that depend on the original reaction rates according to
\begin{eqnarray}
  &&k_1 = \frac{\mu (\sigma - 2 \lambda)}{2 (\sigma + \lambda)} = \mu \epsilon \ ,
  \qquad k_2 = \frac{\sqrt{3} \, \mu \sigma}{2 (\sigma + \lambda)} \ , \nonumber \\
  &&k_3 = \frac{(\sigma - 2 \lambda) (\sigma + \lambda) (11 \sigma^3 + 21 \sigma^2
  \lambda + 3 \sigma \lambda^2 + 2 \lambda^3)}{12 \mu (2 \sigma - \lambda) 
  (7 \sigma^2 - \sigma \lambda + \lambda^2)} \ , \\
  &&k_4 = \frac{\sqrt{3} \, \sigma (5 \sigma^3 - 3 \sigma^2 \lambda + 15 \sigma 
  \lambda^2 - 4 \lambda^3)}{(\sigma - 2 \lambda) (11 \sigma^3 + 21 \sigma^2 \lambda
  + 3 \sigma \lambda^2 + 2 \lambda^3)} \ . \nonumber
\end{eqnarray}
These coefficients encode information about the spatio-temporal pattern formation 
present in this system. Its linear instability is apparent for $k_1 > 0$. The 
resulting oscillatory or spiral instability is saturated by the coefficient of the 
nonlinear term $k_3 > 0$. The associated stochastic noise terms convey information 
about the intrinsic fluctuations in the system and are described in the subsequent
section~\ref{noise}. 

In general, the fields $z_1$ and $z_2$ are complex-valued, and the two independent
Langevin equations \eref{gclecomp} hence contain twice as many degrees of freedom 
as a single dynamical equation for one complex field. For small fluctuations near
the species coexistence fixed point, one may assume the deviations from $\bar{a}$
to be constrained to the real axis; eqs.~\ref{gclecomp} then precisely match the
partial differential equations for the real and imaginary parts of the CGLE 
complex order parameter field, respectively. The connection to the noisy complex 
Ginzburg--Landau equation (CGLE) is thus borne out upon constructing the two linear
combinations $\phi = z_1 + i z_2$ and $\chi = z_1 - i z_2$, or 
$z_1 = (\phi + \chi) / 2$, $z_2 = (\phi - \chi) / 2 i$, with 
$z_1^2 + z_2^2 = \phi \chi$. These obey the Langevin equations
\begin{eqnarray}
  \frac{\partial \phi(\vec{x},t)}{\partial t} &=& D \nabla^2 \phi(\vec{x},t) 
  + (k_1 - i k_2) \, \phi(\vec{x},t) - k_3 (1 - i k_4) \, \phi(\vec{x},t)^2 \,
  \chi(\vec{x},t) \nonumber \\
  &&+ \xi(\vec{x},t) \ , 
\label{cgle1} \\
  \frac{\partial \chi(\vec{x},t)}{\partial t} &=& D \nabla^2 \chi(\vec{x},t) 
  + (k_1 + i k_2) \, \chi(\vec{x},t) - k_3 (1 + i k_4) \, \chi(\vec{x},t)^2 \,
  \phi(\vec{x},t) \nonumber \\
  &&+ \xi'(\vec{x},t) \ .
\label{cgle2}  
\end{eqnarray}
For small and real fluctuations $z_1$, $z_2$, obviously $\chi = \phi^*$, and
eq.~\eref{cgle1} turns into the desired CGLE,
\begin{eqnarray}
  \frac{\partial \phi(\vec{x},t)}{\partial t} &=& D \nabla^2 \phi(\vec{x},t) 
  + (k_1 - i k_2) \, \phi(\vec{x},t) - k_3 (1 - i k_4) \, |\phi(\vec{x},t)|^2 \,
  \phi(\vec{x},t) \nonumber \\
  &&+ \xi(\vec{x},t) \ ,
\label{cgleq}  
\end{eqnarray}
while \eref{cgle2} is merely its complex conjugate.

\section{Noise covariance matrix calculation} \label{noise}

An analysis of the fully stochastic system enables us to systematically account for
internal reaction noise in the system. The noise correlation matrix obtained from 
the stochastic partial differential equations through this path integral approach 
is modified during the course of our derivation of the CGLE. Here we describe the 
steps that lead to the additive noise contributions in the final CGLE \eref{cgleq}.
\footnote{Tracking the full noise correlation matrix in the course of all 
 intermediate steps is rather cumbersome. We have employed Mathematica to aid us 
 with detailed book-keeping and algebraic simplifications.} 
\begin{itemize}
\item 
The noise correlation matrix $L$ written for the fields $a$ is modified into the 
matrix $B$ expressed in terms of the $b$ field variables which are just deviations 
from the mean density $\bar{a}$, 
\begin{equation}
  \langle \zeta'_\alpha \, \zeta'_\beta \rangle = 2 B_{\alpha \beta} \, 
  \delta(\vec{x}-\vec{x\,}') \delta(t-t') \ , \quad 
  B_{\alpha \beta} = L_{\alpha \beta}[a_\alpha \to \bar{a} + b_\alpha] \ .
\end{equation}
\item 
The rotation-like dynamical variable transformation outlined in section~\ref{dvt} 
modifies the correlation matrix as follows,
\begin{equation}\label{rotate}
  \langle \eta_\alpha \, \eta_\beta \rangle = 2 \tilde{B}_{\alpha \beta} \,
  \delta(\vec{x}-\vec{x\,}') \delta(t-t') \ , \quad \tilde{B}_{\alpha \beta} 
  = \Big[ \mathcal{R} B \mathcal{R}^T \Big]_{\alpha \beta} \ .
\end{equation}
\item
In section~\ref{drim}, we simply use the substitution \eref{Kvalue} in $\tilde{B}$.
We note that the resulting noise contributions for the fast field are purely 
multiplicative and of the order $c_1^2$, $c_2^2$; there are no constant terms,
independent of the fluctuating fields. Dimensional reduction of the matrix is 
withheld until the last step, see below.
\item
Our final transformation is the non-linear one outlined in section~\ref{norm}, 
where we use eqs.~\eref{ntz1}, \eref{ntz2} in the matrix obtained in the previous 
step. To zeroth order in the fluctuating fields its entries are constants:
\begin{equation} \label{Bnoise}
\tilde{B}=\left(
\begin{array}{ccc}
  3 \mu ^2 \sigma / 2 (\sigma + \lambda)^2 & 0 & 0 \\
  0 & 3 \mu ^2 \sigma / 2 (\sigma + \lambda)^2 & 0 \\ 0 & 0 & 0 \\
\end{array}
\right).
\end{equation}
As a consequence of rotational symmetry, this matrix is diagonal. 
Note that to this order the dynamical variable transformation \eref{rotate} does 
not `rotate' the noise correlation matrix, and hence generates no constant additive 
term for the fast field, rendering it kinetics deterministic. 
\item
We note that in general, one would need to eliminate the fast degree of freedom in 
the correlation matrix to obtain the conditioned noise 
correlator~\cite{1478-3975-9-6-066002}, 
$\tilde{B'} = \tilde{B}_{22} - \tilde{B}_{23} \tilde{B}_{33} \tilde{B}_{32}$; here, 
$\tilde{B}_{22} = \left(
\begin{array}{cc} 
  3 \mu ^2 \sigma / 2 (\sigma + \lambda)^2 & 0 \\
  0 & 3 \mu ^2 \sigma / 2 (\sigma + \lambda)^2
\end{array} \right)$, $\tilde{B}_{23}^T = \tilde{B}_{32} = (0 \;\ 0)$,  
$\tilde{B}_{33} = 0$.
\item 
To lowest order in the field fluctuations, the final noise correlator $\tilde{B}'$
is simply diagonal and constant
\begin{equation}
  \tilde{B}' = \left( \begin{array}{cc}
  3 \mu ^2 \sigma / 2 (\sigma + \lambda)^2 & 0 \\
  0 & 3 \mu ^2 \sigma / 2 (\sigma + \lambda)^2 
  \end{array} \right)
\end{equation}
\end{itemize} 
and hence describe mere additive noise in the stochastic CGLE \eref{cgleq}.

We emphasize that the assumption of small amplitude fluctuations near the Hopf
bifurcation enables us to justify keeping only the zeroth-order constant terms in 
the noise correlators. We note that for larger fluctuations, additional 
multiplicative noise terms would come into play and indeed become dominant near
the absorbing extinction fixed points. As stochastic trajectories reach the
heteroclinic cycles in the system, the effective description in terms of the CGLE
is thus rendered invalid. As seen in \eref{Bnoise}, for small field fluctuations
the noise in the fast relaxational eigendirection is decoupled from the stochastic 
dynamics on the slow reactive manifold; there are no cross-correlations between the
fast relaxing mode and the slow damped oscillatory modes.

\section{Results and conclusions} \label{result}

In the spatially extended stochastic May--Leonard model variant under consideration 
here, a Hopf bifurcation separates two regimes: (i) For $\sigma < 2 \lambda$, there 
exists a stable limit cycle, and all dynamical degrees of freedom relax towards the
stationary three-species coexistence fixed point. The lattice system is 
correspondingly characterized by finite species patches with typical size $\xi_c$. 
(ii) For $\sigma > 2 \lambda$, in contrast the deterministic limit cycles become
unstable, inducing an instability of a spatially homogeneous state against the
spontaneous formation of spiral structures. Analyzing the associated coefficients of 
the CGLE \eref{cgleq}, we note that this parameter region corresponds to the real 
part $k_1 > 0$ of the coefficient of the linear term. The imaginary part $k_2$ is 
always positive. The growth of the spiral spatio-temporal patterns is ultimately
inhibited by the non-linear terms in the deterministic dynamics. For stable spirals, 
the real part $k_3$ of the coefficient of the non-linear term must be greater than 
zero, which again implies $\sigma > 2 \lambda$, consistent with the instability 
condition for the limit cycle. We note that for small fluctuations in the dynamical
fields, the noise components of the CGLE in the three-species coexistence region are 
merely additive to the lowest order. Hence, at least in this regime we may draw a 
correspondence between the original microscopic reaction rates of the model and the 
coefficients of the effective continuum evolution equation, thus quantitatively
describing spiral pattern formation in the coexistence phase. 

In forthcoming work, we plan to verify our results through detailed Monte Carlo 
simulations for stochastic May--Leonard models on a two-dimensional lattice. We 
hope to map the continuum theoretical parameters to corresponding rates on a 
discrete lattice. We furthermore intend to corroborate the change in pattern 
formation from patches to spirals as we move across the Hopf bifurcation in
parameter space, i.e., from $\sigma < 2 \lambda $ to $\sigma > 2 \lambda $.

To summarize, we demonstrate that a stochastic treatment of the fluctuations due 
to internal reaction noise is possible for the May--Leonard model through the
field-theoretic formalism. We derive a fully stochastic set of partial differential
equations \eref{1ceq}-\eref{3ceq} that incorporates the intrinsic stochasticity of 
the system in the continuum limit. Specifically near the Hopf bifurcation, i.e.,
for $0 < \epsilon \ll 1$, we may exploit the emerging time scale separation and 
eliminate one fast relaxing mode, which at least for small amplitude fluctuations 
leads us to a stochastic complex Ginzburg--Landau equation as a coarse-grained 
dynamical description. We derive the most relevant non-trivial noise effect terms
for this effective CGLE and see that to the lowest order in the fluctuating fields,
additive noise dominates, and cross-correlations between from the fast and slow 
degrees of freedom are absent.
 
We emphasize that for generic parameter values away from the Hopf bifurcation, one 
cannot achieve a similar dimensional reduction as there is no adequate separation 
of time scales. Indeed, the relaxing degree of freedom will couple to the two  
oscillating modes through non-linear feedback. A mapping of the stochastic
May--Leonard model to the CGLE is thus not in general possible. Also, the validity 
of the non-linear transformation employed in our derivation is constrained to small
fluctuations of the dynamical variables. Higher-order terms neglected in this 
procedure become non-negligible when deviations from the stationary coexistence 
fixed point become appreciable. For example, phase space trajectories could then 
traverse the heteroclinic orbit not captured by the simple CGLE. It is important to
note that the field-theoretic Doi--Peliti formalism and the equivalent Langevin 
description in terms of three dynamical fields remain applicable for arbitrary 
values of the intrinsic rates, and could be utilized for further detailed 
mathematical analysis of the spatially extended stochastic May--Leonard model.

\ack
We would like to acknowledge and thank Darka Labavi\'c, Hildegard Meyer-Ortmanns, 
and Mauro Mobilia for fruitful discussions and helpful suggestions associated with 
this work.

%\section*{References}
%\bibliographystyle{unsrt}
%\bibliography{Popdynam}

\end{document}